\documentclass[conference]{IEEEtran}
\IEEEoverridecommandlockouts
\usepackage{cite}
\usepackage{amsmath,amssymb,amsfonts}
\usepackage{algorithmic}
\usepackage{graphicx}
\usepackage{makecell}
\usepackage{textcomp}
\usepackage[utf8]{inputenc} 
\usepackage{booktabs}
\usepackage{hyperref}
\usepackage{tabularx}
\usepackage{xcolor}
\usepackage{soul}

\usepackage[inline]{enumitem}
\usepackage{comment}
\newcolumntype{L}{>{\raggedleft\arraybackslash}X}
\newcolumntype{R}{>{\raggedright\arraybackslash}X}

\def\BibTeX{{\rm B\kern-.05em{\sc i\kern-.025em b}\kern-.08em
    T\kern-.1667em\lower.7ex\hbox{E}\kern-.125emX}}
\begin{document}

\title{Challenges and Governance Solutions for Data Science Services based on Open Data and APIs
}

\author{\IEEEauthorblockN{Juha-Pekka Joutsenlahti and Timo Lehtonen}
\IEEEauthorblockA{\textit{Solita Ltd, Finland}\\
first.last@solita.fi
}
\and
\IEEEauthorblockN{Mikko Raatikainen, Elina Kettunen  and Tommi Mikkonen}
\IEEEauthorblockA{\textit{University of Helsinki, Finland} \\
\textit{first.last@helsinki.fi}}
}

\maketitle

\begin{abstract}

Increasingly common open data and open application programming interfaces (APIs) together with the progress of data science -- such as artificial intelligence (AI) and especially machine learning (ML) --  create opportunities to build novel services by combining data from different sources. In this experience report, we describe our firsthand experiences on open data and in the domain of marine traffic in Finland and Sweden and identified technological opportunities for novel services. We enumerate five challenges that we have encountered with the application of open data: relevant data, historical data, licensing, runtime quality, and API evolution. These challenges affect both business model and technical implementation. We discuss how these  challenges could be alleviated by better governance practices for provided open APIs and data.

\end{abstract}

\begin{IEEEkeywords}
Open data, application programming interface, API, artificial intelligence, machine learning, governance.
\end{IEEEkeywords}

\section{Introduction}
Especially governmental organizations and agencies provide different types of open data sources in several countries today. While not new, open data has not always been available via convenient Application Programming Interfaces (APIs) as the data can also be provided only as documents \cite{Gonzalez2020model}. However, for example in Finland, this will be resolved because the recent Finnish legislation demands governmental organizations to provide APIs for their public data adhering to the European Union directive \cite{EUdirective}. 
The same legislation, thus, increases the amount of available open data and open APIs  as different data will be opened. Additionally, this enables more near real-time data when data will become available automatically through the APIs. 

As the quantity of data and data sources grow massively, a need and opportunity emerge for data science services that can process huge amounts of data. Data can be seen as a service proposal per se, but with open data, the commercial product is rather the intelligence built on top of the data to solve the specific contextual problems of the customers. More and more often, this takes place in the form of artificial intelligence (AI) and machine learning (ML) systems.

However, provided APIs essentially define and control what operations can be performed on open data, by whom, and under what conditions.  
For open data usage, there are challenges, such as legal or privacy issues and possible changes in governmental policies~\cite{Martin2013risk}.
Challenges are also associated with the creation and evolution of data science services, such as ensuring adequate efficiency in processing large data sets (data-intensive flows management)~\cite{Abello2017data}.

This experience report discusses the following problem: How to build data science services on top of a set of open data providers by combining open data sources and carrying out advanced analyses, such as machine learning so that the results are valuable for end-customers? Our firsthand experiences originate from the development of open data and APIs for marine traffic. This open data in association with other open data sources introduce new sustainable data science service opportunities for end-customers based on various AI technologies. Towards this end, we enumerate the challenges we have identified related to these opportunities and propose how API governance could alleviate the challenges.

\section{Case: Artificial intelligence for the Finnish–Swedish Winter Navigation System}

\subsection{Background: Marine traffic}

We base our firsthand experiences on marine traffic data in Finland and Sweden, especially during wintertime when the Finnish–Swedish Winter Navigation System (FSWNS) \cite{bergstrom2020simulation} is active. FSWNS maintains safe and efficient year-round navigation with agreements and information systems. The Finnish public authority (Fintraffic) provides near real-time monitoring of traffic through public APIs\footnote{https://www.digitraffic.fi/en/}.
This case was selected because Solita Ltd\footnote{http://www.solita.fi}, a consultancy company with over 1000 employees, developed the DigiTraffic API for Fintraffic and two authors of this paper work full time at Solita. 
Therefore, we are familiar with the technologies, application domain, possible business opportunities, and challenges.

Marine traffic is quintessential for Finland: some 80-90\% of exported and imported goods are carried by sea \cite{tapaninen2020maritime}. In winter navigation, the changing ice conditions \cite{banda2015risk} cause relatively frequently accidents  \cite{goerlandt2017analysis} which may trigger, for instance, oil spills and delays \cite{banda2016risk}. 
The ecosystem related to marine traffic is large as the total number of companies working in the Finnish maritime cluster\footnote{https://shipowners.fi/en/maritime-cluster/} is almost 3000 \cite{tapaninen2020maritime}. 

The marine environment of FSWNS is very special and challenging as it consists of shallow and narrow sea lanes, dense and rocky archipelago, and icy conditions as especially the Bay of Bothnia freezes during the wintertime  \cite{lehtola2019finding}. These conditions do not only require piloting but also ice breaking in  winter when ships may even be guided in a convoy or towed behind an ice breaker. 
The icebreakers assist vessels free of charge to enable fluent foreign trade. However, unexpected delays of even tens of hours are typical\footnote{https://vayla.fi/en/transport-network/waterways/winter-navigation} due to changing ice conditions which then affect the inter-modal logistic chain \cite{tapaninen2020maritime}. Nowadays, the icebreaker captains try to interpret an ice forecast\footnote{http://baltice.org/weather/}, a wind forecast\footnote{https://www.windy.com} and satellite images to predict which vessels might get stuck into the ice and need assistance. 

\subsection{Opportunities}

Novel data science service utilizing  AI technologies could provide help for the decision making process for the icebreaker captains, but also several other parties in the ecosystem would gain benefits from better information: ports, shipping companies, cargo forwarding companies, and transport companies \cite{tapaninen2020maritime} -- and their customers.
Business-critical opportunities include how to predict and control the estimated time of arrival (ETA) or departure more accurately based on ice and weather conditions. Also, these opportunities may include identifying potential future traffic bottlenecks --  
waiting does not only mean staying on hold and wasting time at the sea but also that the speed of the vessel could have been lower leading to savings if the waiting was anticipated. Respectively, sudden stop, acceleration, or un-optimized route of a large vessel is equally costly. All these accumulate CO$_2$ emissions.
Many manufacturing and assembly companies also depend on predictive just-in-time import and export in their business processes \cite{tapaninen2020maritime}. The logistics are not limited to marine but include storage, road, and rail.
Finally, there is an opportunity to enhance safety,  such as predict and prevent a collision in a convoy \cite{jussila2018visualising}.
Future opportunities lie in technological advancements. Automation is increasingly important in the ports and heavily automated ports already exist outside Finland (e.g., Hamburg and  Rotterdam). Also, the ships rely more and more on automation and autonomous cargo ships are being developed and tested.


The public authorities have already made several open data APIs available that are published following the aforementioned legislation.  For example, the following APIs are relevant and available:
\begin{enumerate*}
    \item Digitraffic (Ministry of Transport and Communications) traffic APIs including
 marine\footnote{https://www.digitraffic.fi/en/marine-traffic/} and rail\footnote{https://www.digitraffic.fi/en/railway-traffic/};
    \item Finpilot piloting status (government-owned piloting company)\footnote{ https://pilotonline.fi/traffic-info/api};
    \item Finnish Meteorological Institute weather\footnote{https://en.ilmatieteenlaitos.fi/open-data-manual};
    \item many others gathered to the Open data webpage\footnote{https://www.avoindata.fi/en}
\end{enumerate*}

The increasingly popular open data facilitates novel business opportunities to create  data science services for end-customer. 
That is, AI and especially ML can provide different stakeholders with advanced analytics and predictions based on the open data of these open APIs. Business-critical challenges include how to find a paying customer for open data and build a sustainable software ecosystem: The raw data cannot be the product as basically anyone can access it, so the value for a customer must come from, for example, user experience and good analytics in the right place and time.  However, rather than focusing only on the challenges in algorithms and technical solutions, business models, or ecosystems, there are also more general software engineering challenges.

\section{Software Engineering Challenges}

In this section, we elaborate software engineering related challenges that we have encountered whilst considering different data science service opportunities based on intelligence built on top of open data and APIs for the maritime cluster.

\subsubsection{Relevant data}
REST APIs are today the dominant design in data APIs. It is customary that full data is provided through a REST API with no ability to customize what data is returned. This often leads to fetching a large amount of unnecessary data. For example, Maritime traffic API contains very large JSON messages but sometimes only one piece of data, such as the ETA of a vessel's portcall, is actually needed. That is, the data providers have just opened all data without much considering how data would be best usable. When multiple APIs are utilized and each provides much unnecessary data, this makes the development and running of the system inconvenient and unreliable. There are already be technologies that could potentially alleviate this. For example, GraphQL APIs can drastically reduce the sizes of transferred JSON files \cite{Brito2020rest}.

\subsubsection{Historical data}
Historical data, i.e. the data produced and collected over the years, is rarely made available. For example, the Maritime API shows only the current traffic although all data over years has been stored. The data providers might not have considered that someone could find value in the historical data on marine traffic. Alternatively, the historical data can be generalized and provided in a more-coarse grained manner in order not to cause too much load for API, such as only for limited time intervals or limited locations as appears to be for some weather data. However, historical data is required for learning in ML systems. Although a data user can start to store data in order to form a training dataset, storing is slow and inconvenient, and brings forward other challenges, such as licensing below.

\subsubsection{Licensing}
When multiple data sources are involved, different rights become an issue. Unfortunately, often open data and API licenses are even more difficult to manage than those of open source software. For example, common software licenses, such as GPL, BSD or MIT, or content licenses, such as Creative Commons (CC), are not necessarily used for data. Rather, open data providers create their own licenses or do not explicitly mark any license. The licenses can be even hard to find in the API specifications, such as for the pilot data above.
Moreover, when data is collected from different sources, it may be difficult to assess how different licenses are compatible and how the new combined data or solution inferred from data can be licensed and commercialized.

\subsubsection{Runtime quality}
A data science service based on multiple sources that need to be accessed near real-time, emphasizes different runtime quality characteristics, such as reliability and availability. With governmental open data, sudden changes to the open data policies are not perhaps as likely as with other organizations. However, there are no guarantees for dependability or service level agreements (SLAs) at least clearly stated in the data sources. The benefit of data science services often lies in near real-time inference and a discontinuity in source data APIs will immediately affect usefulness. In the worst case, the results can be incorrect rather than unavailable.

\subsubsection{API Evolution} 
APIs evolve over time, and the changes often break the client developers' code (e.g. \cite{Brito2020}). Generally, the most common API breaking changes are due to refactorings \cite{Dig2006}. With open data APIs, the changes have the potential to affect several API users and end-customers -- likewise when multiple APIs are used any of them can change and break the solution. A specific problem in ML-based solutions is their lack of fault tolerance if something changes: an ML system can continue its operations and produce incorrect, drifted results if some of its data sources have changed.
Identifying the most likely changes in open APIs and preparing for them, e.g., by a means of fault tolerance, can help to mitigate potential API evolution problems.  In addition, API evolution can also lead to changes in licensing and technical implementation. This in turn adds yet another layer of complexity in the development process.

\section{A need for governance models}
One unifying factor with all the challenges identified in the previous section is that they are all related, to a certain degree, to the governance of the provided APIs and data. \textit{API governance} is defined as \textit{"a task mainly applied inside an organization, typically aiming at achieving a certain harmonization of APIs in terms of their non-functional properties, best-practices-support, documentation quality or rule compliance in general"} \cite{Haupt2018api}. In \cite{De2017api}, API governance is seen encompassing a wide range of activities \textit{"starting with the API proposal all the way to its adoption, through requirements gathering, build and deploy, and operations during general availability"}.
 Data governance, in turn, is associated with decisions regarding the data, i.e., \textit{"data governance refers to who holds the decision rights and is held accountable for an organization’s decision-making about its data assets"}~\cite{Khatri2010designing}. 
API governance encompasses practices that need to be designed and executed to overcome the challenges of building intelligent data science services on top of open data APIs. As a summary, following aspects needs to be considered when designing API governance \cite{horkoff2019strategic}, \cite{krintz2013developing}, \cite{lourencco2019framework}:

    \subsubsection{Change Control} When API changes are required, the effects of the change should be predictable and implemented in a uniform, consistent way. If changes need to be rolled back, the return to previous functionality should also be consistent, complete, and managed. This requires the development of efficient and automated change-impact analysis techniques that can determine the potential effects of a proposed change.
    
    \subsubsection{Impact of Changes} As APIs are created in the context of business, the impact of changes in API should be carefully evaluated. Stakeholders of an API, such as consumers and business owners, should be informed of changes and the possible impact of those.

    \subsubsection{Policy Specification and Analysis} Access control policies, their analysis, and application should be considered to only allow authorized clients to access resources. The accessibility to an API should also consider the business context.

    \subsubsection{Consistent Policy Implementation} Policies that control the use of assets through API should be implemented independently of the technologies that are used to implement the assets. Decoupling API from asset implementation allows for API integrity to be kept changes to one do not influence the other.
    
    \subsubsection{Life-cycle Alignment} The governance process should be involved in all the duration of the API life-cycle. A governance process should exist for the development, deployment, monitoring, and deprecation of an API.
    
    \subsubsection{API Integrity} API should be able to interface on a newer version of the platform without conflicts and without effort. When planning new features, existing API should not require extensive refactoring, and backward compatibility should be ensured over a period of time.
    
    \subsubsection{Monitoring and Auditing} API governance must incorporate a unified method of monitoring and auditing API activity.

\begin{table}[b]
\begin{center}
\caption{Relating API governance aspects and recognized challenges.} \label{tab:summary}
\begin{tabularx}{\columnwidth}{X | p{.25cm} | p{.25cm} | p{.25cm} | p{.25cm} | p{.25cm} | p{.25cm} | p{.25cm}|} 
API governance aspect $\backslash$  challenge &   \rotatebox{90}{Relevant data} &
\rotatebox{90}{Historical data} &
\rotatebox{90}{Licensing} & 
\rotatebox{90}{Runtime quality} &
\rotatebox{90}{Evolution}    \\
\hline 
Change control   & &X &  & X& X  \\ 
\hline
Impact of changes   & X & &  & X& X \\
\hline
Policy specification and analysis    & &   & X & &  \\
\hline
Consistent policy implementation   & &  & X& &  \\
\hline
Life-cycle alignment   & X & &  &X & X  \\
\hline
API integrity   & &  & X &X & X \\
\hline
Monitoring and auditing && X& & X & X \\
\hline \end{tabularx}
\end{center}
\end{table} 

Table \ref{tab:summary} summarizes the cross-mapping of API governance aspects and recognized challenges, which helps to understand how API governance could be designed in order to tackle the recognized challenges. To summarize, to guarantee that only \textit{relevant data} is published through API, there is a need to understand the development life-cycle, manage change and understand how changes impact data consumers. Moreover, creating an API that provides \textit{historical data} requires change control as the data structure behind the API can change, but API still needs to provide the same data, whereas  \textit{licensing} requires policies to be designed and implemented to manage access to API. Also, API integrity needs to be considered with licensing to manage the compatibility and licenses of different API versions. \textit{Runtime quality} requires quality from both data and the API. API integrity, change control, and acknowledging the impact of changes are major factors to ensure runtime quality. Runtime quality can also be improved by API monitoring and auditing. As the need for runtime quality exists during the whole lifespan of an API, life-cycle alignment is required. Finally, successful \textit{evolution} of API requires managing changes from technical and business perspectives, making sure API integrity exists, aligning API life-cycle, and monitoring and auditing the API.

Moreover, technical API governance and data governance share some responsibilities when it comes to data quality management. As traditional APIs, such as REST and SOAP, are usually built on a separate layer that is not directly connected with data \cite{soni2019api}, most of the data quality dimensions, e.g. completeness, interpretability, accessibility, and representational consistency are shared responsibilities between data and API governance. In addition, as API governance also contains technical aspects, it needs to be considered as part of the IT governance as well. 

API governance can be seen as part of the broader practise of API management, that is described as: 
\textit{"An activity that enables organizations to design, publish and deploy their APIs for (external) developers to consume. API Management capabilities such as controlling API lifecycles, access and authentication to APIs, monitoring, throttling, and analyzing API usage, as well as providing security and documentation are often implemented through an integrated platform, which is supported by an API gateway."} \cite{mathijssen2020identification}

API governance can also be seen as a combination of data and IT governance, and as a part of broader API management. There is a need for a governance model that describes the structure of how API governance should be designed and executed. This structure should take into account the different aspects of API governance but enable organizational flexibility that exists because of the heterogeneous nature of business domains.

\section{Conclusions}

Open data and APIs are increasingly important and available in the future. Many different business opportunities can arise based on data science services, especially relying on machine learning built on top of open data and APIs. However, besides finding a viable business model and algorithms, there are several software engineering challenges when combining open data from several different open APIs for dependable data science services. We outlined our firsthand experiences about the challenges that we have encountered whilst working in the domain of maritime traffic and its open data and APIs. The challenges are not all domain-specific but pertain to data science services built on top of open data and APIs. As a solution to challenges, we discussed how better governance practices at open data and API providers could alleviate these challenges for those who design and operate data science services. We aim to carry out more research on data sciences services built on open data and APIs to gain experiences of fully exploiting the potential in a fully dependable manner. Especially, we are interested in different governance practices in the entire value network.

\bibliographystyle{IEEEtran}
\bibliography{bib}

\end{document}